\documentclass[a4paper, 12pt]{article}
\usepackage{a4}
\usepackage{amsfonts}
\usepackage{amssymb}
\usepackage{cite}
\usepackage{epsfig}
\usepackage{graphicx}
\usepackage{psfig}

\author{}

\newcommand{\be}{\begin{equation}}
\newcommand{\ee}{\end{equation}}
\newcommand{\bea}{\begin{eqnarray}}
\newcommand{\eea}{\end{eqnarray}}

\newcommand{\half}{\frac{1}{2}}
\newcommand{\1}{{\bf 1}}
\newcommand{\3}{{\bf 3}}
\newcommand{\2}{{\bf 2}}
\newcommand{\4}{{\bf 4}}
\newcommand{\5}{{\bf 5}}
\newcommand{\6}{{\bf 6}}
\newcommand{\8}{{\bf 8}}

\newcommand{\ov}{\overline}
\newcommand{\tr}{{\rm tr}}

\newcommand{\R}{\mathbb{R}}
\newcommand{\Anti}{{\bf Anti}}
\newcommand{\Sym}{{\bf Sym}}
\newcommand{\Adj}{{\bf Adj}}

\newcommand{\N}{{\bf N}}

\newcommand{\F}{{\bf F}}
\newcommand{\ch}{{\rm ch}}
\def\IR{\relax{\rm I\kern-.18em R}}

\def\IP{\relax{\rm I\kern-.18em P}}
\def\inbar{\vrule height1.5ex width.4pt depth0pt}
\def\IC{\relax\,\hbox{$\inbar\kern-.3em{\rm C}$}}

\def\K3{{\bf K3}}

\def\ov{\overline}

\begin{document}

\title{
\begin{flushright} \vspace{-2cm}
{\small MPP-2006-12 \\
\vspace{-0.35cm}
hep-th/0602101} \end{flushright}
\vspace{4.0cm}
Massive U(1)s and Heterotic Five-Branes on K3
}
\vspace{1.5cm}

\author{\small  Gabriele Honecker }

\date{}

\maketitle
\begin{center}
\emph{Max-Planck-Institut f\"ur Physik, F\"ohringer Ring 6, \\
  80805 M\"unchen, Germany } \\
\vspace{0.2cm}
\tt{gabriele@mppmu.mpg.de}
\vspace{1.0cm}
\end{center}
\vspace{1.0cm}

\begin{abstract}
\noindent  
We systematically consider heterotic $SO(32)$ and $E_8 \times E_8$ compactifications 
on $K3$ with Abelian and non-Abelian backgrounds as well as an arbitrary number of five-branes.
The masses of the $U(1)$ factors depend on the first Chern classes of the bundles and some combinatorial 
factors specifying the embedding in $SO(32)$ or $E_8$. 
The form of the generalised Green-Schwarz counter-terms in six dimensions
constrains the possible heterotic five-brane actions.\\
Some supersymmetric examples on $K3$ realisations as toric complete intersection spaces 
with up to three explicit two-forms are given.

\end{abstract}

\thispagestyle{empty}
\clearpage

\tableofcontents

\section{Introduction}

Recently, it was shown that not only in four-dimensional Type II string compactifications, e.g.  with 
intersecting or magnetised branes (for a recent review see~\cite{Blumenhagen:2005mu}), 
but also in heterotic compactifications, multiple anomalous $U(1)$ factors can 
arise~\cite{Blumenhagen:2005ga,Blumenhagen:2005pm,Blumenhagen:2005zg,Weigand:2005ng}.\footnote{Further four-dimensional 
heterotic compactifications with $U(N)$ bundles can be found in~\cite{Distler:1987ee,Sharpe:1996xn,Andreas:2004ja}.}

In six-dimensional heterotic compactifications this phenomenon has been known for a long time~\cite{Green:1984bx,Strominger:1986uh},
and the $E_8 \times E_8$ case has been investigated in some detail motivated by F-theory, 
see e.g.~\cite{Aldazabal:1996du,Berglund:1998va,Aspinwall:2005qw}. However, 
a fully quantitative treatment of the
Green-Schwarz counter-terms including non-perturbative effects is still missing.
In~\cite{Sagnotti:1992qw}, the role of additional tensor 
multiplets for the generalised Green-Schwarz mechanism was advertised, which is relevant for 
heterotic five (H5)-branes in $E_8 \times E_8$ compactifications.
On the other hand, six-dimensional $SO(32)$ heterotic compactifications with multiple $U(1)$s and H5-banes
have only  been poorly investigated.\footnote{Early six dimensional heterotic string spectra with H5-branes 
- not discussing the detailed contributions of these to the anomaly cancellation - 
and purely non-Abelian gauge groups are given in~\cite{Schwarz:1995zw,Duff:1996rs,Seiberg:1996vs,Aspinwall:1996nk}, 
for more references see also the review~\cite{Ibanez:1997td}.}

In this article, we systematically study six-dimensional $SO(32)$ and $E_8 \times E_8$
heterotic string compactifications on $K3$ with an arbitrary number of
H5-branes extended along the non-compact directions and multiple $U(1)$ gauge factors. 
We show that the masses of the Abelian gauge factors depend on the first Chern classes of the bundles and some
combinatorial factors associated to the embedding of the $U(1)$s in $SO(32)$ or $E_8$, while the instanton numbers, i.e.
second Chern characters, enter the tadpole cancellation condition. 
Moreover, we explicitly compute the H5-brane contributions to the generalised Green-Schwarz counter-terms and show that they 
do not contribute to the Abelian mass terms.
We hope that the present work can help to clarify the F-theory lift of multiple heterotic $U(1)$ factors.

The paper is organised as follows:
After some general remarks on ${\cal N}=1$ heterotic
compactifiactions on $K3$ and anomaly cancellation in six dimensions in section~\ref{S_general},
the $SO(32)$ case is treated in full generality in section~\ref{S_SO32}, and two examples with non-trivial unitary bundles
are given.
The generic $E_8 \times E_8$ case is treated in section~\ref{S_E8}, a class of embeddings with $U(n) \times U(m)$ backgrounds
is specified and three examples are 
discussed exhibiting the relations among first Chern classes and massive $U(1)$ factors. 
Finally, the conclusions are given in section~\ref{S_con}.


\section{Six-dimensional heterotic compactifications}
\label{S_general}

\subsection{Some facts in six dimensions}
\label{Ss_diverse}

In this section, some facts about ${\cal N}=1$ heterotic compactifications on $K3$ are collected.
\begin{itemize}
\item
The ${\cal N}=1$ supersymmetric multiplets in six dimensions are 
the hyper, vector, tensor and supergravity multiplets with field content given in table~\ref{Tab:6Dmultiplets}.
\begin{table}[htb]
\renewcommand{\arraystretch}{1.2}
\begin{center}
\begin{tabular}{|c||c|c|}
\hline
\hline
Multiplet & Content \\\hline
SUGRA & $(g_{\mu\nu}, B_{\mu\nu}^+,\psi_{\mu}^{-})$ 
\\
Tensor & $(B_{\mu\nu}^-,\phi,\chi^{+})$
\\
Vector & $(A_{\mu},\lambda^{-})$
\\
Hyper & $(4\varphi,\psi^+)$
\\\hline
\end{tabular}
\end{center}
\caption{Bosonic and fermionic content of the ${\cal N}=1$ multiplets in six dimensions. The index -(+) denotes a spinor of
negative (positive) chirality or an (anti)selfdual two-form. Half-hyper multiplets can occur if they transform under some 
real representation of the gauge group.}
\label{Tab:6Dmultiplets}
\end{table}
\item
For ${\cal N}=1$ heterotic string compactifications,
the net number of chiral states transforming under some bundle $V$ on a Calabi-Yau $n$-fold is given by
the Riemann-Roch-Hirzebruch theorem~\cite{Nakahara:1990th} 
\bea
\chi(V) = \int_{CY_n} {\rm ch}(V) {\rm Td}(CY_n) 
\label{E_gen_chirality}
\eea
associated to the cohomology classes $H^{\ast}(CY_n,V)$.
The non-trivial Todd classes on $K3$ are given by
\bea
{\rm Td}_0(K3) = 1, \quad {\rm Td}_2(K3) =\frac{1}{12}c_2(K3)=2,
\nonumber 
\eea
leading to  
\bea
\chi_{K3}(V) = {\rm ch}_2 (V) +2r 
\label{E_chi}
\eea
with $r = {\rm rank}(V)$. Since the vector and hyper multiplets in six dimensions have opposite chirality, 
with the sign convention in~(\ref{E_gen_chirality}), the index counts
\bea
\chi_{K3}(V) =  \# {\rm Vector}- \# {\rm Hyper}
\nonumber
\eea
multiplets in the representation associated to the bundle $V$. The gauge group in string compactifications is known by constructions.
Thus, in contrast to four-dimensional compactifications, the complete massless spectrum can be computed from the 
index~(\ref{E_chi}).
\footnote{To be more precise, the bosonic index~(\ref{E_gen_chirality}) is by supersymmetry equal to the fermionic 
Atiyah-Singer index $\int_{CY_n} {\rm ch}(V) \hat{A}(CY_n)$.
The fermionic content in table~\ref{Tab:6Dmultiplets} consist of a pair of 
symplectic Majorana spinors per multiplet. A hyper multiplet contains a complex scalar and symplectic Majorana fermion in 
the {\it complex} representation  $\bf{R}$ as well as the CPT conjugate states in $\ov{\bf R}$, i.e. {\it one} hyper multiplet is denoted
by ${\bf R}+c.c.$ and the number of such multiplets is counted by the index  $\chi(V)$ of the associated bundle, 
with $\chi(V)=\chi(V^{\ast})$  
 on $K3$. For a {\it pseudo-real} representation, a symplectic Majorana spinor is CPT invariant 
and half-hyper multiplets 
can occur which are taken into account in this article by allowing for half integer numbers in table~\ref{Tab:SO32spectrum}
.}
\item
In general, several types of six-dimensional field theory anomalies occur which can be encoded in the well known anomaly eight-form
\bea
I_8 &=& \frac{n_H - n_V + 29 n_T- 273}{360}\tr R^4 +\frac{n_H - n_V -7 n_T+51}{288}(\tr R^2)^2
\label{E:I_polynomial}
\\
&+& \frac{1}{6} \tr R^2 \sum_A C_A \tr F^2_A
-\frac{2}{3}\sum_A A_A \tr F^4_{A} -\frac{2}{3}\sum_A B_A (\tr F^2_A)^2 
\nonumber \\
&+& 4\sum_{A<B} C_{AB} \tr F^2_A \tr F^2_B
+ {8 \over 3}\sum_{A,B} D_{AB} \tr F_A \tr F_B^3,
\nonumber
\eea
where $A_A$, $ B_A$, $C_A$, $C_{AB}$  and $D_{AB}$ are  coefficients encoding the number of fermions transforming under 
some gauge group(s), 
the sign given by their chirality and multiplicities from various representations taken into account. 
The traces formally run also over Abelian gauge factors.
The complete list of coefficients including multiple Abelian factors explicitly can be found, 
e.g., in~\cite{Erler:1993zy}.

As we will show in section~\ref{Ss_GSM}, factorizable anomalies can be cancelled by Green-Schwarz counter-terms whereas the 
$\tr R^4$ and the non-Abelian $\tr F^4$ anomalies have to be absent for a consistent six-dimensional massless spectrum.
\item
The supersymmetry condition, the so called Donaldson-Uhlenbeck-Yau equation (DUY), on some 
{\it holomorphic } background gauge field strength $\ov{F}$ 
in six dimensions is given by
\bea
\int_{K3} J \wedge \ov{F} = 0, 
\label{E:DUY_K3}
\eea
where $J$ is the K\"ahler form on $K3$. Potential loop corrections with the same parity symmetry under $\ov{F} \rightarrow -\ov{F}$
would involve $\ov{F}^{2n+1}$ and must be absent for dimensional reasons. It is thus expected that~(\ref{E:DUY_K3}) is
perturbatively exact.
Further support for this conjecture stems from the fact that the ten-dimensional dilaton forms the scalar degree of freedom of 
the six-dimensional universal tensor multiplet. As we will show, this tensor multiplet contributes to the generalised Green-Schwarz
mechanism, but does not acquire a mass. Any loop correction to~(\ref{E:DUY_K3}) would be in contradiction to this observation.
\item
The Bianchi identity on the three-form field strength results in the so called tadpole cancellation condition  given by
\bea
\tr \ov{F}^2 - \tr \ov{R}^2 -16 \pi^2 N_{H5} =  0 
\label{E:gen_tcc} 
\eea
in cohomology on $K3$,
where $N_{H5}$ is the total number of H5-branes for both $SO(32)$ and $E_8 \times E_8$ string compactifications.
\item
A gauge field $F=d A$ in $D$ dimensions becomes massive through a coupling~\cite{Douglas:1996sw,Berkooz:1996iz} 
\bea
m \int_{\R^{1,D-1}} \gamma^{(D-2)} \wedge F \sim  m \int_{\R^{1,D-1}} \left(\star_D d \beta^{(0)} \right) \wedge A,
\label{E:explanation_mass}
\eea
with the duality relation $d \gamma^{(D-2)} \sim \star_D d \beta^{(0)} $ among the $(D-2)$ form $\gamma$ and scalar $\beta$.
The coupling on the left hand side naturally arises in the dimensional field theory reduction presented in section~\ref{Ss_GSM}
while the right hand side has the more familiar shape~\cite{Green:1984bx}.\\
The massive gauge factor remains as a perturbative global symmetry in the Lagrangian.\\
Heterotic compactifications contain at most 16 massive $U(1)$ factors from the perturbative gauge group.
\item
Consistent models are further constrained by K-theory. In~\cite{Witten:1985mj,Freed:1986zx}
it has been shown that at least for compactifications to four dimensions, the K-theory constraint is given by
\bea
c_1(V_{total}) = 0 \, {\rm mod} \, 2,
\label{E:Ktheory}
\eea
where $V_{total}$ is the total background gauge bundle.
Since K-theory is associated to ${\mathbb Z}_2$ valued charges, we expect the condition~(\ref{E:Ktheory}) to hold also for 
$K3$ compactifications.
\item
If supersymmetry is preserved, a vector multiplet becomes massive by absorbing a complete hyper multiplet. 
The coupling~(\ref{E:explanation_mass})
absorbs one scalar descending from the ten-dimensional antisymmetric tensor and the DUY equation~(\ref{E:DUY_K3})
freezes one geometric modulus.
The freezing of the remaining two scalars in a hyper multiplet is best seen in the ${\cal N}=2$, $d=4$ language:
a hyper multiplet contains a scalar triplet of $SU(2)_R$, for which three Fayet-Iliopoulos terms arise, see e.g.~\cite{Douglas:1996sw}.
In order to preserve supersymmetry, all three $D$-terms have to vanish simultaneously.
In the present case, the triplet consists of geometric moduli, and one $D$-term condition provides the DUY equation~(\ref{E:DUY_K3}), 
while the other two supersymmetry conditions are encoded in the requirement of a holomorphic vector bundle. 
The two geometric moduli which would deform the vector bundle from a pure $(1,1)$-form to contain a $(2,0)$ or $(0,2)$ piece are frozen.
\end{itemize}


\subsection{K3 toy models as complete intersection spaces}
\label{S:K3_Examples}

There exist three simple  possibilities to express $K3$ as complete intersections of projective spaces
without introducing singularities.
The three possibilities including the Quartic are as follows,
\bea
{\cal M}_1 = \IP_3 [4], \quad\quad
 {\cal M}_2=\matrix{ \IP_1 \cr  \IP_2 \cr}\hskip -0.1cm\left[{\matrix{ 2\cr 3\cr}}\right] , \quad\quad
 {\cal M}_3=\matrix{ \IP_1 \cr  \IP_1 \cr \IP_1 \cr}\hskip -0.1cm\left[{\matrix{ 2\cr 2\cr 2\cr}}\right],
\nonumber
\eea
with in the toric description up to three explicit two-forms.

Let $\eta_i$ denote the $(1,1)$-forms on the up to three projective factors. The Stanley-Reisner ideals are 
then given by
\bea
SR^{(1)} = \{ \eta^4\}, \quad\quad 
SR^{(2)} = \{\eta_1^2, \eta_2^3\}, \quad\quad  
SR^{(3)} = \{\eta_1^2,\eta_2^2, \eta_3^2 \},   
\nonumber
\eea
leading to the intersection forms
\bea
I_2^{(1)} = 4 \eta^2,  \quad\quad 
I_2^{(2)} = 3 \eta_1 \eta_2 + 2 \eta_2^2,  \quad\quad  
I_2^{(3)} = 2 \eta_1 \eta_2 +  2 \eta_1 \eta_3 +  2 \eta_2 \eta_3. 
\label{E:IntersectionForms}
\eea
It can be checked explicitly that all three manifolds have complex dimension two and 
\bea
c_1(T{\cal M}_i)=0, \quad\quad c_2(T{\cal M}_i)=24, \quad i=1,2,3.
\eea

The conditions for the parameter space of a model to lie inside the K\"ahler cone are as follows,
\bea
\int_{K3} J \wedge \eta_i \stackrel{!}{>} 0 \quad {\rm for \; all \; } i, 
\quad\quad \int_{K3} J \wedge J  \stackrel{!}{>} 0,
\eea
where $J$ is the K\"ahler form on ${\cal M}_i$.\footnote{In the formulation of the full $(3,19)$ lattice,
these conditions are replaced by the K\"ahler form being selfdual.}
Expanding the K\"ahler form in terms of the $(1,1)$-forms ($\ell_s \equiv 2 \pi \sqrt{\alpha'}$)
\bea
J = \ell_s^2 \sum_i \rho_i \eta_i
\nonumber
\eea
gives the conditions for the three $K3$ realisations presented here
\bea
&(1)&  \rho \stackrel{!}{>} 0,
\nonumber\\
&(2)& \rho_2  \stackrel{!}{>} 0, 
\quad\quad  3 \rho_1 + \rho_2\stackrel{!}{>} 0,    \nonumber\\ 
&(3)& \rho_i+\rho_j \stackrel{!}{>} 0 {\rm \; for \; } i \neq j, 
\quad\quad \rho_1\rho_2 +\rho_1\rho_3+\rho_2\rho_3 \stackrel{!}{>} 0.  
\nonumber
\eea
Abelian bundles are specified completely by their first Chern classes, \mbox{$c_1(L) = \sum_i q_i \eta_i$}, 
and the DUY equations for the different cases are computed using the corresponding intersection forms~(\ref{E:IntersectionForms}),
\bea
&(1)& q \rho \stackrel{!}{=} 0,
\nonumber\\
&(2)& 3q_1\rho_2+ 3q_2\rho_1+2q_2\rho_2 \stackrel{!}{=} 0,
\label{E:DUYs_Ps}
\\
&(3)& q_1[\rho_2+\rho_3] +q_2[\rho_1+\rho_3]  +q_3[\rho_1+\rho_2]  \stackrel{!}{=} 0.
\nonumber
\eea

If $K3$ is realised as the Quartic ${\cal M}_1=\IP_3[4]$, non-trivial line bundles cannot solve the DUY equation, 
and only models with pure $SU(n)$ bundles can preserve supersymmetry. Moreover, line bundles are specified 
(up to a sign)
by their second Chern characters due to $\ch_2(L)=2 q^2$.
The situation is different for the two remaining $K3$ realisations ${\cal M}_2$ and ${\cal M}_3$,  as we will show in some examples in 
sections~\ref{Ss:SO32_models_Ex_1},~\ref{Ss:SO32_models_Ex_2} and~\ref{Ss:E8_models_Exs}.

A  vector bundle $V$ of rank $r$ on a complete intersection manifold
${\cal M}$ can be defined by the cohomology of the monad as $V={\rm Ker}(f)/{\rm Im}(g)$,
\bea
0 \rightarrow {\cal O}|_{\cal M}^{\oplus p} \stackrel{g}{\rightarrow} \oplus_{a=1}^{r+p+q}{\cal O}(n_1^a,\ldots,n_k^a)\vert_{\cal M} 
\stackrel{f}{\rightarrow}  \oplus_{b=1}^{q} {\cal O}(m_1^b,\ldots,m_k^b)\vert_{\cal M} \rightarrow 0,
\nonumber
\eea
with $p \geq 0, q  \geq 1$ and $k$ the number of $(1,1)$-forms on $\cal{M}$ 
(for more details on the notation see~\cite{Blumenhagen:2005ga}), 
or alternatively via the exact sequence 
\bea
0 \rightarrow V  \rightarrow \oplus_{a=1}^{r+q}{\cal O}(n_1^a,\ldots,n_k^a)\vert_{\cal M} 
\stackrel{f}{\rightarrow}  \oplus_{b=1}^{q} {\cal O}(m_1^b,\ldots,m_k^b)\vert_{\cal M} \rightarrow 0.
\nonumber
\eea
In both cases, the Chern classes are computed from
\bea
c(V)=\frac{\prod_a\left(1+\sum_i n_i^a \eta_i\right)}{\prod_b\left(1+\sum_i m_i^b \eta_i\right)},
\nonumber
\eea
leading in particular to
\bea
c_1(V) &=& \sum_i \left(\sum_a n_i^a - \sum_b m_i^b \right) \eta_i,
\nonumber\\
\ch_2(V) &=& \half \sum_i \left(\sum_a (n_i^a)^2 - \sum_b (m_i^b)^2  \right) \eta_i^2
+\sum_{i<j}  \left(\sum_a n_i^a n_i^a - \sum_b m_i^b m_j^b\right) \eta_i \eta_j.
\nonumber
\eea
A necessary condition for a well-defined stable bundle $V$ is $n_i^a,m_i^b ,m_i^b-n_i^a \geq 0$ for all $i,a,b$ and 
$(m_1^b-n_1^a,\ldots, m_k^b-n_k^a) \neq (0,\ldots,0)$. Stability is guaranteed if all defining maps $f$ and $g$ have maximal rank.
We will, however, not check the latter condition explicitly for the examples given in this article.

The DUY equations for a vector bundle $V$ 
are given by~(\ref{E:DUYs_Ps}) when replacing $q_i \rightarrow \left(\sum_a n_i^a - \sum_b m_i^b \right)$.


\subsection{The perturbative Green-Schwarz counter-terms}
\label{Ss_GSM}

If some general bundle is embedded into $SO(32)$ or $E_8 \times E_8$,
several types of anomalies involving gauge and gravitational fields can
occur:
while $\tr R^4$ and $\tr F^4$ field theory anomalies in six dimensions have to be absent,
factorizable anomalies can in general be cancelled by a generalized
Green-Schwarz mechanism~\cite{Green:1984sg,Green:1984bx}. In contrast to four dimensions, where only mixed and pure Abelian
anomalies are compensated, in six dimensions also anomalies involving only
gravity and non-Abelian gauge fields  have counter-terms.

In the following, we perform the dimensional reduction of the heterotic string
on $K3$ similar to the reduction on Calabi-Yau three-folds
treated in~\cite{Blumenhagen:2005ga,Blumenhagen:2005pm}. 
The relevant couplings linear in the antisymmetric tensor $B^{(2)}$ and its ten-dimensional dual $B^{(6)}$ 
arise from the kinetic and the one-loop counter-term in ten dimensions,\footnote{The prefactor of the Green-Schwarz counter-term has 
been derived in~\cite{Blumenhagen:2006ux} from M-theory reduction and by S-duality.}
\bea
S_{kin} &=& -\frac{\pi}{\ell_s^8} \int_{\R^{1,9}} e^{-2\phi_{10}}\, H\wedge \star_{10}\, H,
\nonumber\\
S_{1-loop} &=&  {1\over 24\, (2\pi)^3\, \ell_s^2}\,  \int_{\R^{1,9}} B^{(2)}\wedge X_8,
\eea
with the field strength $H^{(3)}= dB^{(2)}-{\alpha'\over 4}(\omega_{Y}-\omega_{L})$.
The anomaly eight-form is given by~\cite{Green:1984sg}
\bea
  X_8={1\over 24} {\rm Tr} F^4 -{1\over 7200} \left( {\rm Tr} F^2\right)^2 
      -{1\over 240} \left( {\rm Tr} F^2\right) \left( {\rm tr} R^2\right)+
       {1\over 8}{\rm tr} R^4 +{1\over 32} \left( {\rm tr} R^2\right)^2
\eea
for both the  $SO(32)$ and $E_8 \times E_8$ theories.
Denoting by $X_{n+\ov{m}}$ a form with $n$ legs along $\R^{1,5}$ and $m$ legs
along $K3$, the terms relevant for the Green-Schwarz mechanism take the
form
\bea
S_{kin} &=& {1 \over 8\pi \ell_s^6} \int_{\R^{1,5} \times K3}
\left(\left[\tr \ov{F}^2-\tr \ov{R}^2  \right]\wedge B^{(6)} 
+\left[\tr F^2-\tr R^2  \right]\wedge B^{(\bar{4}+2)}
+ 2 \, \tr (F \ov{F})\wedge B^{(\bar{2}+4)} 
\right),
\nonumber\\
S_{1-loop} &=& {1 \over 24 (2\pi)^3 \ell_s^2} \int_{\R^{1,5} \times K3}
\left(B^{(\bar{2})}\wedge X_{\bar{2}+6} +B^{(2)}\wedge X_{\bar{4}+4} \right).
\eea
Expanding in terms of a basis $\{\omega_k\}_{k=0,\ldots,h_{11}+1}$ of two-forms on $K3$
as well as its dual basis $\{\widehat{\omega}_l\}_{l=0,\ldots,h_{11}+1}$, i.e. 
$\int_{K3} \omega_k \wedge \widehat{\omega}_l = \delta_{kl}$,\footnote{The following expansion 
of $\bar{Y}$ differs by a factor of $2\pi$ from the one in~\cite{Blumenhagen:2005ga,Blumenhagen:2005pm}.} 
\bea
B^{(2)} &=& b^{(2)}_0 + \ell_s^2 \sum_{k=0}^{h_{11}+1} b^{(0)}_k \omega_k, \quad\quad
B^{(6)} = c^{(6)}_0 + \ell_s^2 \sum_{k=0}^{h_{11}+1}
c^{(4)}_k \widehat{\omega}_k + \ell_s^4 c^{(2)}_0 {\rm vol}_4,
\nonumber\\
\bar{Y}_{\ov{2}} &=&  \sum_{k=0}^{h_{11}+1} \left[\bar{Y}\right]^k \omega_k
=  \sum_{k=0}^{h_{11}+1} \left[\bar{Y}\right]^{\widehat{k}} \widehat{\omega}_k,
\label{E:B_exansion}
\eea
with $\int_{K3}{\rm vol}_4 =1$ and $k=0,h_{11}+1$ labeling the (2,0) and
(0,2) form, respectively, leads to the six-dimensional couplings linear in $b^{(i)}_k$ and $c^{(i)}_k$
\bea
S_{kin} &=& {1 \over 8\pi \ell_s^6} \int_{\R^{1,5}} c^{(6)}_0\int_{ K3}
\left[\tr \ov{F}^2-\tr \ov{R}^2  \right]
\nonumber\\
&&+{1 \over 8\pi \ell_s^2}\int_{\R^{1,5}}  c^{(2)}_0 \wedge \left[\tr F^2-\tr R^2  \right]
\nonumber\\
&&+{1 \over 4 \pi \ell_s^4} \sum_{k=0}^{h_{11}+1} \int_{\R^{1,5}}c^{(4)}_k \wedge
\left[\tr (F\ov{F}) \right]^{k},
\nonumber\\
S_{1-loop} &=& {1 \over 24 (2\pi)^3} \sum_{k=0}^{h_{11}+1} \int_{\R^{1,5}}
 b^{(0)}_k \, \left[X_{\bar{2}+6}\right]^{\widehat{k}}
\nonumber\\
&&+{1 \over 24 (2\pi)^3 \ell_s^2} \int_{\R^{1,5}}
b^{(2)}_0\wedge\int_{ K3} X_{\bar{4}+4}.
\label{E_dim_red}
\eea
The scalars $ b^{(0)}_k$ (or their dual four-forms $c^{(4)}_k$) belong in six dimensions
to the 20 hyper multiplets encoding the $K3$ geometry. In more detail, 19 hyper multiplets contain
one scalar $b^{(0)}_k$ pertaining to the anti-selfdual two-forms on $K3$ and three geometric moduli each, 
while one hyper multiplet contains 
the remaining three $ b^{(0)}_k$ belonging to the three selfdual two-forms on $K3$ and the overall volume modulus.
The two-forms  $b^{(2)}_0$ (or its duals $c^{(2)}_0$) contain the 
selfdual tensor $B_{\mu\nu}^+$ of the six-dimensional supergravity multiplet and the 
anti-selfdual tensor $B_{\mu\nu}^-$ of the universal tensor multiplet. The scalar degree of freedom in the 
universal tensor multiplet is given by the dilaton.

\begin{figure}
\begin{center}
\includegraphics[width=0.9\textwidth]{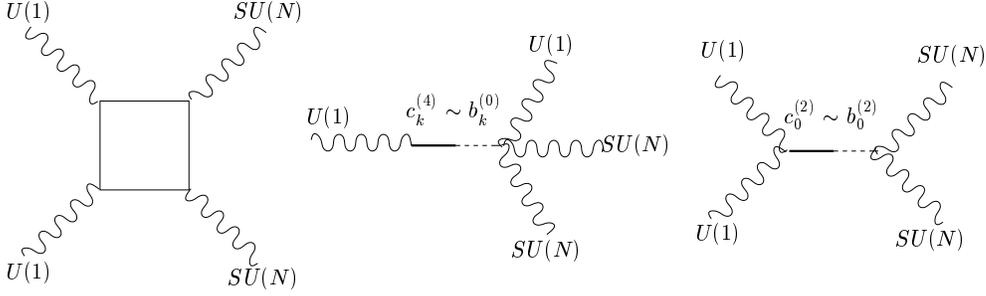}
\end{center}
\caption{A six-dimensional anomalous diagram and its two possible types of Green-Schwarz counter diagrams.}
\label{Fig:GS}
\end{figure}

The terms~(\ref{E_dim_red}) combine to two types of Green-Schwarz counter-terms~\cite{Berkooz:1996iz} depicted 
in figure~\ref{Fig:GS},
\bea
{\cal I}_{pert} &=& \frac{1}{48 (2 \pi \ell_s)^4}\int_{K3}\left(
\tr (F \ov F) \wedge X_{\ov{2}+6} + \frac{1}{2}\left(\tr F^2 - \tr R^2 \right)\wedge  X_{\ov{4}+4}
\right),
\label{E_GS_pert}
\eea
where the first term corresponds to the sum over all counter diagrams of the first type 
with $c^{(4)}_k \sim b^{(0)}_k$ exchange\footnote{These have been missed in~\cite{Green:1984bx}.}
and the last term corresponds to the second Green-Schwarz counter diagram involving $c^{(2)}_0 \sim b^{(2)}_0$ couplings.
As in four dimensions, massive $U(1)$ factors occur if some coupling 
\bea
S_{mass} = {1 \over 4 \pi \ell_s^4} \sum_{k=0}^{h_{11}+1} \int_{\R^{1,5}}c^{(4)}_k \wedge \left[\tr (F\ov{F})\right]^{k}
\label{E:U1mass}
\eea
exists, but the Abelian factor can still be anomaly-free if the (sum of the) Green-Schwarz diagrams vanish(es). On the other hand, contrarily to
the four-dimensional case, $U(1)^4$ and $U(1)^2-\tr R^2$ anomalies can be cancelled by diagrams involving only 
$b^{(2)}_0 \sim c^{(2)}_0$ exchange 
in the same way as $(\tr_{SU(N)}F^2)^2$ and $\tr_{SU(N)}F^2-\tr R^2$ counter-terms exist. An anomalous $U(1)$ factor in six 
dimensions can hence stay massless and a massive $U(1)$ factor can be anomaly-free. 
These effects are important for the correct identification of the F-theory lift.

Note that a supersymmetric background $\ov{F}$ is of type $(1,1)$ and can thus only receive a mass via couplings
to $c^{(4)}_k$ for $k=1,\ldots, h_{11}$. This will be the case in the examples discussed in sections~\ref{Ss:SO32_models_Ex_1},~\ref{Ss:SO32_models_Ex_2}
and~\ref{Ss:E8_models_Exs}.

Except from the perturbative terms~(\ref{E_GS_pert}), H5-branes  contribute to the anomaly cancellation.
While for $SO(32)$ compactifications, they provide symplectic gauge factors as well as charged matter, in the $E_8 \times E_8$ theory
they provide one tensor and one hyper multiplet per H5-brane.

The detailed shape of the polynomials  $X_{\ov{2}+6}$, $X_{\ov{4}+4}$ as well as the H5-brane contributions is discussed
in sections~\ref{S_SO32} and~\ref{S_E8} for $SO(32)$ and $E_8 \times E_8$ compactifications, respectively.


\section{The $SO(32)$ case}
\label{S_SO32}

\subsection{Massless spectrum, H5-branes and tadpole cancellation}
\label{Ss_SO32_H5tcc}

For $SO(32)$ heterotic compactifications, we decompose
\bea
SO(32) \supset SO(2M) \times \prod_{i=1}^{K} U(n_i \, N_i)
\nonumber
\eea
with $M + \sum_{i=1}^{K}n_i \, N_i = 16$ and take bundles 
\bea
W = \oplus_{i=1}^{K} V_i
\label{E:SO32_bundles}
\eea
with structure group $G = \prod_{i=1}^{K} U(n_i)$ which leave the non-Abelian
gauge group
\bea
H=  SO(2M) \times \prod_{i=1}^{K} SU( N_i)
\nonumber
\eea
as well as the massless $U(1)$ factors to be computed below in the low energy effective field theory.
This prescription
generalises the embedding of $U(1)$ factors in $SO(32)$ discussed in~\cite{Green:1984bx} to $U(n_i)$ bundles with $n_i \geq 1$. 

$N_a$ H5-branes extended along the non-compact dimensions and at the same point $a$ in $K3$ support the gauge group $Sp(2N_a)$ and
have the Chern-Simons couplings~\cite{Blumenhagen:2005zg}
\bea
S_{H5}^{SO(32)} &=& - \frac{2\pi N_a}{\ell_s^6}\int_{\mathbb{R}^{1,5}} B^{(6)} +\frac{1}{4\pi \ell_s^2}
\int_{\mathbb{R}_{1,5}} B^{(2)} \wedge \left(\frac{N_a}{24}\tr R^2 -\tr_{Sp(2N_a)} F^2 \right),
\eea
which were derived by S-duality~\cite{Polchinski:1995df} arguments and are consistent with four-dimensional anomaly cancellation.

Introducing the skyscraper sheafs ${\cal O}\vert_{a}$ with
\bea
{\rm ch} \, ({\cal O}\vert_{a}) = (0,0,-1)
\nonumber
\eea
allows to generalise the index~(\ref{E_chi}) to extensions Ext$^{\ast}_{K3}$.
The massless spectrum is thus computed along the same lines as for the four dimensional case 
in~\cite{Blumenhagen:2005pm,Blumenhagen:2005zg} by decomposing the adjoint of $SO(32)$ and identifying the bundles.
The complete six-dimensional massless spectrum is listed in table~\ref{Tab:SO32spectrum}.\footnote{Note that although 
$\chi({\cal O}\vert_{a},{\cal O}\vert_{a})=0$, there exists a vector multiplet in the 
symmetric(=adjoint) and a hyper multiplet in the antisymmetric representation of $Sp(2N_a)$ in the spectrum.}
\begin{table}[htb]
\renewcommand{\arraystretch}{1.2}
\begin{center}
\begin{tabular}{|c||c|c|}
\hline
\hline
reps. & \# Hyper  & \# Vector  \\
\hline \hline
$(\Adj_{U(N_i)})_{0(i)}$ & $1-\half\chi(V_i \otimes V_i^{\ast})$   & 1\\
\hline
$(\Sym_{U(N_i)})_{2(i)}+c.c.$ &  $-\chi(\wedge^2 V_i)$ & 0 \\
$(\Anti_{U(N_i)})_{2(i)}+c.c.$ & $-\chi(\otimes^2_s V_i)$ & 0\\
\hline

$(\N_i,\N_j)_{1(i),1(j)}+c.c.$ & $-\chi(V_i \otimes V_j)$  & 0    \\
$(\N_i,\ov \N_j)_{1(i),-1(j)}+c.c.$ & $-\chi(V_i \otimes V_j^{\ast})$   & 0  \\
\hline\hline
$(\Adj_{SO(2M)})$ &  0 & 1   \\
$(2{\bf M}, \N_i)_{1(i)}+c.c.$ &  -$\chi(V_i)$ & 0   \\
\hline\hline
$(\Sym_{Sp(2N_a)})$  & 0 & 1  \\
$(\Anti_{Sp(2N_a)})$ &   1 & 0   \\
\hline 
$(\N_i,2\N_a)_{1(i)}+c.c.$ &  $n_i$ & 0  \\
$(2{\bf M}, 2\N_a)$ & $\half$ & 0 
\\\hline
\end{tabular}
\end{center}
\caption{Six-dimensional charged spectrum for the $SO(32)$ heterotic string with bundles of type~(\ref{E:SO32_bundles}) and H5-branes. The
 spectrum is completed by the supergravity and one universal tensor multiplet as well as 
20 neutral hyper multiplets encoding the $K3$ geometry.}
\label{Tab:SO32spectrum}
\end{table}

The tadpole cancellation condition for bundles of type~(\ref{E:SO32_bundles})
is on $K3$ given by
\bea
\sum_{i=1}^{K} N_i \; {\rm ch}_2(V_i)  - \sum_{a=1}^L N_a  = -c_2(T) = -24,
\label{E:SO32_tcc}
\eea
where $\sum_{a=1}^L N_a=N_{H5}$ is the total number of H5-branes.

In total, we have a vanishing $\tr R^4$ anomaly due to 
\bea
n_T &=& 1,
 \nonumber\\
\quad n_H-n_V &=& 
20 - \sum_i \frac{N_i}{2} \left(N_i \chi(V_i \otimes V_i^{\ast})+ (N_i+1)\chi(\wedge^2 V_i)+(N_i-1)\chi(\otimes^2_s V_i)\right)
 \nonumber\\
&&-\sum_{i<j}N_iN_j\left(\chi(V_i \otimes V_j)+\chi(V_i \otimes V_j^{\ast})\right)
-M(2M-1)
 \nonumber\\
&&
-2M \sum_i N_i \chi(V_i) -N_a(2N_a+1)+N_a(2N_a-1)+2N_a \bigl(\sum_i n_i N_i +  M \bigr)
 \nonumber\\
&=&
244,
\eea
where in the last line the tadpole cancellation condition~(\ref{E:SO32_tcc}) and the properties of Chern characters, e.g.
$\ch(V \otimes W) = \ch (V) \cdot \ch(W)$, have been used.\footnote{The index~(\ref{E_chi}) on $K3$ contains only {\it even}
Chern characters in contrast to $CY_3$ compactifications where only {\it odd} ones appear. The relations
among the first Chern characters of $\wedge^2 V$ and the rank $r$ bundle $V$ are e.g. listed 
in eq. (19) in~\cite{Blumenhagen:2005ga}, and by using the relation $V \otimes V =(\wedge^2 V) \oplus (\otimes^2_s V)$ one 
obtains $\ch (\otimes^2_s V) = \frac{r(r+1)}{2} + (r+1) c_1(V) + \left[(r+2)\ch_2(V) +\half c_1(V)^2 \right]+ \cdots$.}

This leads together with table~\ref{Tab:SO32spectrum} to the anomaly eight-form
\bea
I^{SO(32)}_8 & = &\left(\tr R^2 \right)^2
+ \tr R^2 \left(\tr_{SO(2M)} F^2 -2 \, \tr_{Sp(2N_a)} F^2 \right)
\nonumber\\
&&+ \tr R^2 \left( 
\sum_i 2 \left( n_i + 2 \ch_2 (V_i) \right) \tr_{U(N_i)} F^2+\frac{1}{3} \left( \sum_{i}c_1(V_i) \, \tr_{U(N_i)} F\right)^2 \right)
\nonumber\\
&& - \frac{16}{3}\sum_{i,j}c_1(V_i)c_1(V_j) \, \tr_{U(N_i)} F \,  \tr_{U(N_j)} F^3 
\nonumber\\
&&-2 \left(\tr_{SO(2M)} F^2 \right)^2
-8 \left( \sum_i n_i \tr_{U(N_i)} F^2\right)\left( \sum_j (\ch_2(V_j)+n_j) \tr_{U(N_j)} F^2\right)
\nonumber\\
&&+2 \, \tr_{Sp(2N_a)} F^2 \, \tr_{SO(2M)} F^2
+4 \, \tr_{Sp(2N_a)} F^2 \, \sum_i n_i \tr_{U(N_i)} F^2 
\nonumber\\ 
&& -4 \, \tr_{SO(2M)} F^2 \, \sum_i \tr_{U(N_i)} F^2 \left(\ch_2(V_i) + 2 n_i \right),
\nonumber
\eea
which can be rewritten in the partially factorised form
\bea
I^{SO(32)}_8&=&
\left(\tr R^2 -\tr_{SO(2M)} F^2-2 \, \sum_i n_i \, \tr_{U(N_i)} F^2 \right)
\times
\label{E:SO32_AnomalyPolynomial}
\\ 
&&\times
\left(\tr R^2 +2 \, \tr_{SO(2M)} F^2 + 4\,\sum_j (\ch_2(V_j)+n_j) \tr_{U(N_j)} F^2 -2 \, \tr_{Sp(2N_a)} F^2
\right)
\nonumber\\ 
&&+\frac{1}{3} \left( \sum_{i}c_1(V_i) \, \tr_{U(N_i)} F \right)
\times
\left( \sum_{j}c_1(V_j) \left[\tr R^2\,\tr_{U(N_j)} F  - 16 \, \tr_{U(N_j)} F^3 
\right]
\right)
,
\nonumber
\eea
whose non-Abelian part is in agreement e.g. with the  special cases $SO(32) \times Sp(48)$~\cite{Schwarz:1995zw} 
and $SO(28) \times SU(2)$ for $(N,n,\ch_2(V))=(2,1,-12)$~\cite{Erler:1993zy,Duff:1996rs}.

It can be checked explicitly that all non-Abelian $\tr F^4$ anomalies vanish upon tadpole cancellation,
e.g.
\bea
I_{\tr_{U(N_i)} F^4} &\sim& N_i \chi(V_i \otimes V_i^{\ast})+ (N_i+8)\chi(\wedge^2 V_i)+(N_i-8)\chi(\otimes^2_s V_i)
\nonumber\\
&&
+\sum_{j \neq i} N_j\left(\chi(V_i \otimes V_j)+\chi(V_i \otimes V_j^{\ast})\right)
+2 M \chi(V_i) + 2N_a n_i
= 0
,
\nonumber
\eea
for the spectrum in table~\ref{Tab:SO32spectrum}.

In the following section we show that the Green-Schwarz counter-terms have exactly the correct shape to cancel all 
the anomalies encoded in the polynomial~(\ref{E:SO32_AnomalyPolynomial}).


\subsection{Anomaly cancellation for $SO(32)$}
\label{Ss_SO32_GS}

Inserting the expansion~(\ref{E:B_exansion}) in the H5-brane Chern-Simons action gives
\bea
S_{H5}^{SO(32)} 
&=& - \frac{2\pi N_a}{\ell_s^6}\int_{\mathbb{R}^{1,5}} c^{(6)}_0 
-\frac{1}{4\pi \ell_s^2}
\int_{\mathbb{R}_{1,5}} b^{(2)}_0 \wedge \left(  \tr_{Sp(2N_a)} F^2-\frac{N_a}{24}\tr R^2 \right),
\eea
which provides the missing term for the tadpole cancellation condition and
leads to the non-perturbative part of the Green-Schwarz counter-terms,
\bea
{\cal I}_{non-pert} = \frac{1}{192 (2 \pi)^2 \ell_s^4}\left(\tr F^2 - \tr R^2 \right)
\wedge \left( N_{H5} \tr R^2 - 24 \sum_{a=1}^L \tr_{Sp(2N_a)}F^2 \right).
\label{E:SO32_H5GS}
\eea
For the class of bundles~(\ref{E:SO32_bundles}), the relevant polynomials are given by
\bea 
X_{\bar{4}+4} &=&  \sum_{j=1}^K \tr_{U(N_j)} F^2 
 \left(12 \, \tr_{U(n_j)} \ov{F}^2-{n_j \over 4} \, \tr \ov{R}^2 \right)
-{1 \over 8} \tr_{SO(2M)} F^2 \, \tr\ov{R}^2
\nonumber\\
&& + \tr R^2 
\left({1\over 16}  \tr\ov{R}^2 - {1 \over 4}\sum_{j=1}^K N_j \tr_{U(n_j)} \ov{F}^2 \right),
\nonumber\\
X_{\bar{2}+6} &=& \sum_{j=1}^K \tr_{U(n_j)} \ov{F} 
\left( 8 \,\tr_{U(N_j)} F^3 -\half \, \tr_{U(N_j)} F \wedge \tr R^2
\right),
\label{E:SO32_Xs}
\eea
and all other traces  can be extracted from appendix~B in~\cite{Blumenhagen:2005zg}.

From~(\ref{E_GS_pert}),~(\ref{E:SO32_H5GS}) and~ (\ref{E:SO32_Xs}), the complete Green-Schwarz counter-term can be computed,  
\bea
{\cal I}_{pert} + {\cal I}_{non-pert} = -\frac{1}{96 (2 \pi \ell_s)^4} \, I^{SO(32)}_8.
\eea
As required, the counter-terms match (up to some normalisation constant) minus the anomaly eight-form~(\ref{E:SO32_AnomalyPolynomial}).

The masses~(\ref{E:U1mass}) of the Abelian gauge factors
for the class of models presented here are given by
\bea
S^{SO(32)}_{mass} =  \sum_{k=0}^{h_{11}+1} \sum_{i=1}^K 
\frac{M_i^k}{\ell_s^4} \int_{\R^{1,5}}c^{(4)}_k \wedge f_i 
\quad\quad {\rm with } \quad\quad M_i^k = \left[ N_i \, c_1(V_i) \right]^k,
\nonumber
\eea
where $f_i$ is the $U(1)$ part of the $U(N_i)$ gauge factor. The number of massive Abelian gauge factors is given by rank$(M)$.


\subsection{Example 1: $U(3) \times U(3)$ bundle without H5-branes}
\label{Ss:SO32_models_Ex_1}

An example on ${\cal M}_2$ with $U(3) \times U(3)$ bundles is given by
\bea
&& 0 \rightarrow V_1  \rightarrow  {\cal O}(1,0)^{\oplus 2} \oplus {\cal O}(0,1)^{\oplus 2}  \rightarrow {\cal O}(1,3) \rightarrow 0,
\nonumber\\
&& 0 \rightarrow V_2  \rightarrow  {\cal O}(1,0)^{\oplus 2} \oplus {\cal O}(0,1)^{\oplus 2}  \rightarrow {\cal O}(3,1) \rightarrow 0,
\nonumber
\eea
from which the Chern characters are computed,
\bea
c_1(V_1) &=& \eta_1 - \eta_2,  \quad\quad \ch_2(V_1) =-16,
\nonumber\\
c_1(V_2) &=& \eta_2 - \eta_1,  \quad\quad \ch_2(V_2) = -8.
\nonumber
\eea
In particular, both bundles have the same rank but different instanton numbers, i.e. $\ch_2(V_1) \neq \ch_2(V_2)$, 
due to the asymmetric shape of the intersection form~(\ref{E:IntersectionForms}) on ${\cal M}_2$. 

The bundle 
\bea
V= V_1 \oplus V_2
\nonumber
\eea
saturates the tadpole cancellation condition and satisfies the K-theory constraint trivially with $c_1(V)=0$.
The resulting spectrum is listed in table~\ref{Tab:SO32_Ex1}.
\begin{table}[htb]
\renewcommand{\arraystretch}{1.2}
\begin{center}
\begin{tabular}{|c||c|c||c||c|}
\hline
\hline
$SO(20) \times U(1)^2$ & \# H  & \# V & $SO(20) \times U(1)^2$ & \# H    \\
\hline \hline
$(\1)_{0,0}$ & 52 & 2 & $(\1)_{2,0}+c.c.$ & 12 \\
$({\bf 190})_{0,0}$ & 0 & 1 & $(\1)_{0,2}+c.c.$ & 4
\\\hline
 $({\bf 20})_{1,0}+c.c.$ & 10 & 0 & $(\1)_{1,1}+c.c.$ & 50\\
 $({\bf 20})_{0,1}+c.c.$ & 2 & 0 & $(\1)_{1,-1}+c.c.$ & 58
\\\hline
\end{tabular}
\end{center}
\caption{Charged spectrum of example 1 including massive $U(1)$ factors.}
\label{Tab:SO32_Ex1}
\end{table}
Due to $c_1(V_1) = -c_1(V_2) $, the linear combination $U(1)_1 + U(1)_2$ remains massless, while its orthogonal combination
becomes massive by absorbing one neutral hyper multiplet. 

The DUY condition is identical for both bundles, 
\bea
\rho_2=3 \rho_1
\nonumber
\eea
and freezes one K\"ahler modulus as expected from the existence of one massive vector.
It can be easily fulfilled inside the K\"ahler cone.


\subsection{Example 2: $U(3) \times U(3) \times U(1)$ bundle with H5-branes}
\label{Ss:SO32_models_Ex_2}

As a second example, consider the following $U(3) \times U(3) \times U(1)$ bundle 
\bea
V = V_1 \oplus V_2 \oplus L
\nonumber
\eea
on ${\cal M}_3$ defined by
\bea
&& 0 \rightarrow V_1  \rightarrow  {\cal O}(1,0,0)^{\oplus 2} \oplus {\cal O}(0,1,0)^{\oplus 2} \rightarrow {\cal O}(2,1,1) \rightarrow 0,
\label{E:Def_V_bundles_ex2}\\
&&0 \rightarrow V_2  \rightarrow  {\cal O}(0,0,1)^{\oplus 2} \oplus {\cal O}(0,1,0)^{\oplus 2}  \rightarrow {\cal O}(1,2,1) \rightarrow 0,
\nonumber
\eea
with the Chern characters
\bea
c_1(V_1) &=& \eta_2 - \eta_3, \quad\quad \ch_2(V_1) = -10,
\nonumber\\
c_1(V_2) &=& \eta_3 - \eta_1, \quad\quad \ch_2(V_2) = -10,  
\nonumber
\eea
as well as the line bundle
\bea
c_1(L) &=& \eta_1 - \eta_2, \quad\quad  \ch_2(L) = -2.
\nonumber
\eea
The K-theory constraint is again trivially fulfilled with $c_1(V)=0$.
The DUY equations for $V_1$, $V_2$ and $L$ require $\rho_2=\rho_3$, $\rho_1=\rho_3$ and $\rho_1=\rho_2$, respectively.

In order to satisfy the tadpole cancellation condition, two H5-branes are needed.
The charged spectrum for coincident H5-branes is listed in table~\ref{Tab:SO32_Ex2}.
\begin{table}[htb]
\renewcommand{\arraystretch}{1.2}
\begin{center}
\begin{tabular}{|c||c|c||c||c|}
\hline
\hline
$SO(18) \times Sp(4) \times U(1)^3$ & \# H  & \# V & $SO(18) \times Sp(4) \times U(1)^3$ & \# H   \\
\hline \hline
$(\1,\1)_{0,0,0}$ & 40 & 3  &$(\1,\4)_{1,0,0}+c.c.$ & 3  \\
$({\bf 153},\1)_{0,0,0}$ & 0 & 1  & $(\1,\4)_{0,1,0}+c.c.$ & 3 \\
$(\1,{\bf 10})_{0,0,0}$ & 0 & 1 & $(\1,\4)_{0,0,1}+c.c.$ & 1\\
\hline
$({\bf 18},\1)_{1,0,0}+c.c.$ & 4 & 0 & $({\bf 18},\4)_{0,0,0}$ & $\half$\\
$({\bf 18},\1)_{0,1,0}+c.c.$ & 4 & 0 & $(\1,\6)_{0,0,0}$ & 1\\
\hline
$(\1,\1)_{2,0,0}+c.c.$ & 6 & 0  &  $(\1,\1)_{1,0,1}+c.c.$ & 8\\ 
$(\1,\1)_{0,2,0}+c.c.$ & 6 & 0  & $(\1,\1)_{1,0,-1}+c.c.$ & 12\\  
$(\1,\1)_{1,1,0}+c.c.$ & 40 & 0 & $(\1,\1)_{0,1,1}+c.c.$ & 8\\
$(\1,\1)_{1,-1,0}+c.c.$ & 44 & 0 &  $(\1,\1)_{0,1,-1}+c.c.$ & 12
\\\hline
\end{tabular}
\end{center}
\caption{Charged spectrum of example 2 including massive $U(1)$ factors.}
\label{Tab:SO32_Ex2}
\end{table}
The combination $U(1)_1 + U(1)_2 + U(1)_3$ remains massless, while the two orthogonal linear combinations become massive by 
absorbing one hyper multiplet each. This agrees with the fact that $\rho_1=\rho_2=\rho_3$ freezes two K\"ahler moduli.


\section{The $E_8 \times E_8$ case}
\label{S_E8}

\subsection{A specific class of models}
\label{Ss_E8_models}

In order to check the general form of the counter-terms to be derived in sections~\ref{Ss_E8_Gs_perturbative} 
and~\ref{Ss_E8_H5contribution}, we introduce a
series of decompositions of the form \mbox{$E_8^{(i)} \rightarrow E_{r_i} \times SU(n_i+m_i)$} with $i=1,2$,
$r_i+n_i+m_i=9$, and $E_r = E_7, E_6, SO(10), SU(5)$, \mbox{$SU(2) \times SU(3)$} for $r=7,6,5,4,3$. 
Furthermore, we embed bundles with structure group\\ 
\mbox{$SU(n_i) \times SU(m_i) \times U(1)_i$} in $ SU(n_i+m_i)$ by either using  bundles of the form
\bea
V &=& V_1 \oplus V_2,
\label{E:E8bundle_1}
\\
V_i &=& V_{n_i} \oplus V_{m_i} \oplus L_i \quad {\rm with }\quad  c_1(V_{n_i})=c_1(V_{m_i})=0,\; c_1(L_i) \neq 0,
\nonumber
\eea
or
\bea
W &=& W_1 \oplus W_2,
\label{E:E8bundle_2}
\\
W_i &=& W_{n_i} \oplus W_{m_i} \quad {\rm with }\quad  c_1(W_{n_i})=-c_1(W_{m_i}) \neq 0. 
\nonumber
\eea
In the latter case, the total bundle in each $E_8$ factor 
has vanishing first Chern class, $c_1(W_i)=0$, and K-theory does not further constrain the bundles.

The spectrum is computed along the lines described in~\cite{Blumenhagen:2005ga} and gives again the generalisation of\cite{Green:1984bx}
to non-Abelian bundles. 
For example, consider the decomposition $E_8 \rightarrow SU(5) \times SU(5) \rightarrow SU(3) \times SU(2) \times U(1)
\times SU(5)$
and the corresponding breaking
${\bf 248} \rightarrow ({\bf 10},\5) + \ldots \rightarrow (\3,\2;\5)_1+
(\ov{\3},\1;\5)_{-4}+(\1,\1;\5)_6+ \ldots $ from which the bundles associated to the observable $\5$ representations 
with different $U(1)$ charges are read off as $W_3 \otimes W_2$, $\wedge^2 W_3$ and $\wedge^2 W_2$, respectively. 

The general resulting spectrum for bundles of type~(\ref{E:E8bundle_2}) is displayed in tables~\ref{Tab:E8spectrum} 
and~\ref{Tab:E8spectrum_chiral_6d_morecases}.
The multiplicities for bundles of type~(\ref{E:E8bundle_1}) are obtained from the same tables by simply replacing
\bea
W_n = V_n \otimes L^{-m/\mu}, \quad W_m = V_m \otimes L^{n/\mu},
\label{E:E8_VW_relation}
\eea
with $\mu \equiv \gcd(n,m)$.
\begin{table}[htb]
\renewcommand{\arraystretch}{1.2}
\begin{center}
\begin{tabular}{|c|c|c|c|c|c||c||c|c|}
\hline
\hline
$E_7$ & $E_6 $ & $SO(10)$ &  $SO(10)$ & $SU(5)$  & $SU(5)$ & $E_r$ & \# H &\# V \\
\hline \hline
$(1,1)$ & $(2,1)$ & $(3,1)$ & $(2,2)$ & $(4,1)$ & $(3,2)$ & $(n,m)$ & & \\
\hline \hline
$({\bf 133})_0$ & $({\bf 78})_0$ & $({\bf 45})_0$  & $({\bf 45})_0$ & $({\bf 24})_0$  & $({\bf 24})_0$ & $({\bf Adj})_0$ & 0 & 1\\
\hline
$(\1)_0$ & $(\1)_0$ & $(\1)_0$ & $(\1)_0$ & $(\1)_0$ & $(\1)_0$ & $(\1)_0$ & $\!\!2-\half\chi(W_n \otimes W_n^{\ast})\!\!$ & 1 \\
 &  &  &  &  &  &  & $-\half\chi(W_m \otimes W_m^{\ast})$ &  \\\hline
$({\bf 56})_{-1}^{+cc}$ & $({\bf 27})_{-1}^{+cc}$ & $({\bf 16})_{-1}^{+cc}$ & $({\bf 16})_{-1}^{+cc}$ & $(\ov{\bf 10})_{-1}^{+cc}$ & 
 $(\ov{\bf 10})_{-2}^{+cc}$ & $({\bf X})_{\frac{-m}{\mu}}^{+cc}$ & $-\chi(W_n)$ & 0\\
- &  $({\bf 27})_{2}^{+cc}$ & $({\bf 16})_{3}^{+cc}$ & $({\bf 16})_{1}^{+cc}$ & $(\ov{\bf 10})_{4}^{+cc}$ &   
 $(\ov{\bf 10})_{3}^{+cc}$  & $({\bf X})_{\frac{n}{\mu}}^{+cc}$ &  $-\chi(W_m)$ & 0\\ 
\hline
 - &  - & $({\bf 10})_2^{+cc}$ & $\half({\bf 10})_0$ & $(\5)_3^{+cc}$  & $(\5)_1^{+cc}$ & 
$(\F)_{\frac{n-m}{\mu}}^{+cc}$ & $-\chi(W_n \otimes W_m)$ & 0\\
 - &  - & - &  $({\bf 10})_2^{+cc}$ & $(\5)_{-2}^{+cc}$ & $(\5)_{-4}^{+cc}$ &  
$(\F)_{\frac{-2m}{\mu}}^{+cc}$ & $-\chi(\wedge^2 W_n)$ & 0\\  
 - &  - & - & - & - 
& $(\5)_6^{+cc}$ & $(\F)_{\frac{2n}{\mu}}^{+cc}$ & 
$-\chi(\wedge^2 W_m)$ & 0\\  
\hline
$(\1)_{-2} ^{+cc}$ & $(\1)_{-3}^{+cc}$ & $(\1)_{-4}^{+cc}$ & $(\1)_{-2}^{+cc}$ & $(\1)_{-5}^{+cc}$ & $(\1)_{-5}^{+cc}$ 
& $(\1)_{-\frac{n+m}{\mu}}^{+cc}$ 
& $-\chi(W_n \otimes W_m^{\ast})$ & 0 \\
\hline
\end{tabular}
\end{center}
\caption{Part I: Six-dimensional spectrum from the breaking of a single $E_8$ factor. The spectrum is completed by 
the states from the  second $E_8$ factor, the supergravity and one universal tensor multiplet, 
20 neutral hyper multiplets encoding the $K3$ geometry as well as one tensor and neutral hyper multiplet per H5-brane. 
For shortness we abbreviate $\mu \equiv\gcd(n,m)$ and label by ${}^{+cc}$ the complex conjugate representation. }
\label{Tab:E8spectrum}
\end{table}

\begin{table}[htb]
\renewcommand{\arraystretch}{1.2}
\begin{center}
\begin{tabular}{|c|c|c||c||c|c|}
\hline
\hline
&& & $SU(2) \times SU(3)$ & \# H & \# V\\
\hline \hline
$(5,1)$ & $(4,2)$ & $(3,3)$ & $(n,m)$ & & \\
\hline \hline
$(\3,\1)_0$ & $(\3,\1)_0$ &  $(\3,\1)_0$  &  $(\Adj_2,\1)_0$ & 0 & 1 \\
$(\1,\8)_0$ & $(\1,\8)_0$ & $(\1,\8)_0$ &  $(\1,\Adj_3)_0$ & 0 & 1   \\
$(\1,\1)_0$ & $(\1,\1)_0$ & $(\1,\1)_0$ & $(\1,\1)_0$ & $2-\half\chi(W_n \otimes W_n^{\ast})$ & 1\\
&&&& $-\half\chi(W_m \otimes W_m^{\ast})$ &\\
\hline
$(\2,\3)_{-1}^{+cc}$ & $(\2,\3)_{-1}^{+cc}$ & $(\2,\3)_{-1}^{+cc}$ & $(\2,\3)_{-\frac{m}{\mu}}^{+cc}$ & $-\chi(W_n)$ & 0\\
$(\2,\3)_{5}^{+cc}$ & $(\2,\3)_{2}^{+cc}$ & $(\2,\3)_{1}^{+cc}$ & $(\2,\3)_{\frac{n}{\mu}}^{+cc}$  & $-\chi(W_m)$ & 0\\ 
\hline
$(\1,\ov{\3})_{4}^{+cc}$ & $(\1,\ov{\3})_{1}^{+cc}$  & $(\1,\ov{\3})_{0}^{+cc}$ & $(\1,\ov{\3})_{\frac{n-m}{\mu}}^{+cc}$ & 
$-\chi(W_n \otimes W_m)$  & 0\\
$(\1,\ov{\3})_{-2}^{+cc}$ & $(\1,\ov{\3})_{-2}^{+cc}$  & $(\1,\ov{\3})_{-2}^{+cc}$ & $(\1,\ov{\3})_{-\frac{2m}{\mu}}^{+cc}$  &
 $-\chi(\wedge^2 W_n)$  & 0\\
- & $(\1,\ov{\3})_{4}^{+cc}$  & $(\1,\ov{\3})_{2}^{+cc}$ & $(\1,\ov{\3})_{\frac{2n}{\mu}}^{+cc}$  & $-\chi(\wedge^2 W_m)$  & 0
\\  
\hline
$(\1,\1)_{-5}^{+cc}$ & $(\1,\1)_{-3}^{+cc}$ & $(\1,\1)_{-2}^{+cc}$ & $(\1,\1)_{-\frac{n+m}{\mu}}^{+cc}$ & $-\chi(W_n \otimes W
_m^{\ast})$   & 0\\
\hline
 $(\2,\1)_{-3}^{+cc}$ & $(\2,\1)_{-3}^{+cc}$ & $(\2,\1)_{-3}^{+cc}$ & $(\2,\1)_{-\frac{3m}{\mu}}^{+cc}$ 
&  $-\chi(\wedge^3 W_n)$  & 0\\
 -  & $\half(\2,\1)_{0}$ & $(\2,\1)_{-1}^{+cc}$ & $(\2,\1)_{\frac{n-2m}{\mu}}^{+cc}$ & $-\chi((\wedge^2 W_n) 
\otimes W_m)$  & 0\\
\hline
\end{tabular}
\end{center}
\caption{Part II: Six-dimensional spectrum for the decomposition \mbox{$E_8 \rightarrow SU(2) \times SU(3) \times SU(6)$} and 
$U(n) \times U(m)$ bundles embedded in $SU(6)$. }
\label{Tab:E8spectrum_chiral_6d_morecases}
\end{table}

The tadpole cancellation condition for bundles of type~(\ref{E:E8bundle_1}) reads
\bea
\sum_{i=1}^2\left[{\rm ch}_2(V_{n_i}) + {\rm ch}_2(V_{m_i}) +  b_{n_i,m_i} \, \ch_2(L_i) \right] -N_{H5} = -c_2(T)=-24,
\eea
with $b_{n,m} = \frac{nm(n+m)}{(\gcd(n,m))^2}$ and for bundles of type~(\ref{E:E8bundle_2}) it is given by
\bea
\sum_{i=1}^2\left[{\rm ch}_2(W_{n_i}) + {\rm ch}_2(W_{m_i}) \right] -N_{H5} = -c_2(T)=-24.
\eea

The gravitational anomalies are determined by
\bea
n_T=1 +N_{H5}, \quad n_H-n_V =244-29\, N_{H5},
\label{E:CountingStates_E8}
\eea
for an arbitrary $E_8 \times E_8$ heterotic compactification to six dimensions.
The resulting  anomaly polynomial for bundles of type~(\ref{E:E8bundle_2}) reads 
\bea
I^{E_8\times E_8}_8&=& \left[1-\frac{N_{H5}}{8}\right]\left( \tr R^2 \right)^2 
\nonumber\\
&& + \tr R^2 \sum_{i=1}^2 \Bigl[\frac{a_{r_i}}{2}  \tr_{E_{r_i}} F^2 \left({\rm ch}_2(W_{n_i}) + {\rm ch}_2(W_{m_i}) +10 \right)
\nonumber\\
&& \quad\quad\quad\quad\quad
  +b_{n_i,m_i} f^2_i  \left({\rm ch}_2(W_{n_i}) + {\rm ch}_2(W_{m_i}) +a_{n_i,m_i} c_1(W_{n_i})^2  +10 \right)
\Bigr]
\nonumber\\
&& - \sum_{i=1}^2 \Bigl\{ \frac{a_{r_i}^2}{2}\left(\tr_{E_{r_i}} F^2\right)^2\left({\rm ch}_2(W_{n_i}) + {\rm ch}_2(W_{m_i}) +12 \right)
\nonumber\\
&& \quad\quad\quad
+2a_{r_i}b_{n_i,m_i}\, f^2_i \, \tr_{E_{r_i}} F^2 \left({\rm ch}_2(W_{n_i}) + {\rm ch}_2(W_{m_i})
+a_{n_i,m_i} c_1(W_n)^2  +12 \right)
\nonumber\\
&& \quad\quad\quad
+2b_{n_i,m_i}^2\, f^4_i \left({\rm ch}_2(W_{n_i}) + {\rm ch}_2(W_{m_i})+2 a_{n_i,m_i} c_1(W_{n_i})^2  +12 \right)
\Bigr\}.
\eea
The coefficients are defined as follows,
\bea
a_r &=& \frac{1}{6} ,  \frac{1}{4}, 1, 2, (2,2) \quad {\rm for} \quad r=7,6,5,4,(2,1)
\label{E:E8_constants}
\\
a_{n,m} &=& \frac{n+m}{n\, m}, \quad  \quad
b_{n,m} = \frac{n\, m\, (n+m)}{(\gcd(n,m))^2}, \quad  \quad 
\kappa_{n,m} = \frac{n+m}{\gcd(n,m)}, 
\nonumber
\eea
and the polynomial for bundles of type~(\ref{E:E8bundle_1}) is easily obtained by using relation~(\ref{E:E8_VW_relation}).

\subsection{The perturbative Green-Schwarz counter-terms}
\label{Ss_E8_Gs_perturbative}

Before computing the counter-terms for the embeddings~(\ref{E:E8bundle_1}),~(\ref{E:E8bundle_2}) we start by 
rewriting $X_8$ for $E_8 \times E_8$ in full generality. Its relevant components for compactifications to six dimensions are given by
\bea
X_{\bar{4}+4} &=& \left\{\tr F^2_1 \left(
\half \tr \ov{F}^2_1 - {1 \over 4}\tr \ov{F}^2_2  - {1 \over 8} \tr \ov{R}^2 \right)
+ \left[\tr (F_1 \ov{F}_1) \right]^2
+(1 \leftrightarrow 2) \right\}
-\tr (F_1 \ov{F}_1) \tr (F_2 \ov{F}_2)
\nonumber\\
&& +\tr R^2\left({1 \over 16} \tr \ov{R}^2 - {1 \over 8}\tr \ov{F}^2_1 - {1 \over 8}\tr \ov{F}^2_2   \right),
\nonumber\\
X_{\bar{2}+6} &=& \left\{\tr F^2_1 \left( \tr (F_1 \ov{F}_1) - \half \, \tr (F_2 \ov{F}_2)\right)
+(1 \leftrightarrow 2) \right\}
-{1 \over 4} \, \tr R^2  \left( \tr (F_1 \ov{F}_1) + \tr (F_2 \ov{F}_2)\right),
\nonumber
\eea
and after inserting the tadpole cancellation condition~(\ref{E:gen_tcc}), we obtain
\bea
X_{\bar{4}+4} &=& \left\{\frac{3}{4} \, \tr F^2_1 \left(\tr \ov{F}^2_1 - \half\tr \ov{R}^2 \right)
+ \left[\tr (F_1 \ov{F}_1) \right]^2
+(1 \leftrightarrow 2) \right\}
-\tr (F_1 \ov{F}_1) \tr (F_2 \ov{F}_2)
\nonumber\\
&&-{1 \over 16} \tr R^2 \tr \ov{R}^2
-4\pi^2 N_{H5} \left[  \tr F^2_1 + \tr F^2_2 \right] - 2\pi^2 N_{H5}  \tr R^2,
\nonumber
\eea
which can serve as a guidance to the correct contributions from H5-branes to the generalised Green-Schwarz mechanism.

The perturbative contributions to the counter-terms for any gauge background in $E_8 \times E_8$
are therefore given by
\bea
{\cal I}_{pert} &=&   \frac{1}{48 (2 \pi \ell_s)^4}\int_{K3} \Bigl(
\frac{3}{2}\left[\tr F_1^2 \left\{ \tr F_1^2 \left(\frac{1}{4}\tr \ov{F}_1^2  - \frac{1}{8} \tr \ov{R}^2  \right)
+ \left(\tr(F_1\ov{F}_1)\right)^2 \right\}
 + (1 \leftrightarrow 2)\right]
\nonumber\\
&& \quad \quad \quad \quad \quad \quad\quad
-\frac{1}{4}\tr R^2 \left[ \tr F_1^2 \left(\frac{3}{2}\tr \ov{F}_1^2 - \frac{5}{8} \tr \ov{R}^2  \right)
+ 3 \left(\tr(F_1\ov{F}_1)\right)^2 
+ (1 \leftrightarrow 2)\right]
\nonumber\\
&& \quad \quad \quad \quad \quad \quad\quad
+\frac{1}{32} (\tr R^2)^2 \tr \ov{R}^2 \Bigr)
\label{E:E8_Counter_pert}
\\
&& + \frac{N_{H5}}{192 (2 \pi)^2 \ell_s^4} \left[
2 (\tr F_1^2)(\tr F_2^2) -2(\tr F_1^2)^2  -2(\tr F_2^2)^2 
 +  \tr R^2 \, (\tr F_1^2 +\tr F_2^2)  + (\tr R^2)^2 \right].
\nonumber
\eea
The discussion of the H5-brane contributions is postponed to section~\ref{Ss_E8_H5contribution}.

We now proceed to the comparison with the anomaly eight-form for 
the embeddings presented in section~\ref{Ss_E8_models}. The relevant traces for the spectra in tables~\ref{Tab:E8spectrum} 
and~\ref{Tab:E8spectrum_chiral_6d_morecases}
are computed as
\bea
\tr F^2_i &=& a_{r_i} \tr_{E_{r_i}} F^2 + 2\, b_{n_i,m_i} f^2_i, 
\nonumber
\eea
for both kinds of bundles and for bundles of type~(\ref{E:E8bundle_1}) we have
\bea
\tr (F_i\ov{F}_i) &=& 2\, b_{n_i,m_i} f_i\ov{f}_i,
\nonumber\\
\tr \ov{F}^2_i &=& 2  \left(\tr_{SU(n_i)} \ov{F}^2 +\tr_{SU(m_i)} \ov{F}^2 + b_{n_i,m_i} \ov{f}^2_i  \right), 
\nonumber
\eea
with $\ov{f}_i = 2\pi \, c_1(L_i)$,
whereas for the bundle type~(\ref{E:E8bundle_2}) one obtains
\bea
\tr (F_i\ov{F}_i) &=& 2\, \kappa_{n_i,m_i} f_i\ov{f}_i,
\nonumber\\
\tr \ov{F}^2_i &=& 2 \left(\tr_{U(n_i)} \ov{F}^2 +\tr_{U(m_i)} \ov{F}^2 \right),
\nonumber
\eea
with $\ov{f}_i =  2\pi \, c_1(W_{m_i})$.
The constants $a_r, b_{n,m}, \kappa_{n,m}$ have been defined in~(\ref{E:E8_constants}).

It can be checked that all perturbative counter-terms have the correct shape to cancel the anomalies for $N_{H5}=0$.

The mass terms for Abelian gauge fields are 
\bea
S_{mass}^{E_8 \times E_8} &=& \sum_{k=0}^{h_{11}+1}\sum_{i=1}^2 \frac{ M_i^k}{\ell_s^4}  
\int_{\R^{1,5}}c^{(4)}_k \wedge  f_i 
\nonumber
\eea
with 
\bea
 M_i^k = \left[b_{n_i,m_i}\, c_1(L_i) \right]^k, \quad\quad  M_i^k = \left[\kappa_{n_i,m_i}\, c_1(W_{m_i}) \right]^k,
\nonumber
\eea
for bundles of type~(\ref{E:E8bundle_1}) and~(\ref{E:E8bundle_2}), respectively.
The number of massive vectors is given by rank$(M)$.


\subsection{H5-brane contributions to the Green-Schwarz mechanism}
\label{Ss_E8_H5contribution}

Little is known about the field theory of H5-branes in (compactifications of) the ten-dimensional $E_8 \times E_8$ 
theory.\footnote{There are some results starting from the M-theory picture, see e.g.~\cite{Lukas:1998hk}.} 
At this point, we use the anomaly polynomial to find the correct H5-brane contributions to the Green-Schwarz mechanism.

The purely gravitational eight form contribution is given in the first line of~(\ref{E:I_polynomial}) in full generality,
whereas all mixed and pure gauge anomalies depend on the specific embedding. However, to gather some 
information about possible H5-brane contributions, it is sufficient to notice that in no gauge anomaly computation from the spectrum
 the tadpole cancellation condition is used. Therefore, the overall counter-term for mixed and pure
gauge anomalies
must be independent of $N_{H5}$. With the knowledge of the perturbative part~(\ref{E:E8_Counter_pert})
and the counting of multiplets~(\ref{E:CountingStates_E8}), this leads to the expected form
\bea
{\cal I}_{np} = \frac{N_{H5}}{192 (2 \pi)^2 \ell_s^4} \left(2 \,(\tr F_1^2)^2 + 2\,(\tr F_2^2)^2 -2 \,(\tr F_1^2)(\tr F_2^2)
 -  \tr R^2 \, (\tr F_1^2 +\tr F_2^2)  + \half (\tr R^2)^2 \right).
\label{E:E8_non_per_GS}
\eea
 
In~\cite{Sagnotti:1992qw}, it was noticed that the kinetic terms of the six-dimensional tensor fields 
contribute to the generalised Green-Schwarz mechanism. In the present discussion, these are exactly the anti-selfdual tensors
with support on the H5-branes besides the universal tensor multiplet already taken into account in the perturbative
counter-terms in section~\ref{Ss_E8_Gs_perturbative}.

With the ansatz of one anti-selfdual field $d\tilde{b}^{(2)}_s = - \star_6 d\tilde{b}^{(2)}_s$ 
per tensor multiplet ($s=1,\ldots, N_{H5}$),
the corresponding field strength takes the form
\bea
\tilde{H}^{(3)}_s &=& d\tilde{b}^{(2)}_s-\frac{\alpha^{\prime}}{8} \left( a_1 \, \omega_{Y,1} +a_2 \, \omega_{Y,2} - b \, \omega_L \right)
\eea
where the constants $a_i, b$ are not yet specified.
If we take the kinetic term of the H5-brane tensor multiplet with the same normalisation as the one for the universal tensor multiplet 
used in section~\ref{Ss_E8_Gs_perturbative},
\bea
S_{kin} &=& -\frac{\pi}{\ell_s^8} \int_{\mathbb{R}^{1,5} \times K3} d B^{(2)} \wedge d B^{(6)}
= -\frac{\pi}{\ell_s^4} \int_{\mathbb{R}^{1,5}} db^{(2)}_0 \wedge dc^{(2)}_0  + \cdots
\nonumber
\eea
we obtain\footnote{The canonical string frame normalisation of the kinetic term might involve a factor of $g_s^{-2}$. Modifying the
anti-selfduality relation $d\tilde{b}^{(2)}_s = - g_s^{-2} \star_6 d\tilde{b}^{(2)}_s$ accordingly does not change the 
resulting counter-term.} 
\bea
S_{kin,H5}^{E_8 \times E_8} &=& 
 - \frac{\pi}{\ell_s^4} \sum_{s=1}^{N_{H5}} \int_{\mathbb{R}^{1,5}}\tilde{H}^{(3)}_s \wedge \star_6 \tilde{H}^{(3)}_s
\label{E:E8_H5_kin}\\ 
&=& -\frac{\pi}{\ell_s^4} \sum_{s=1}^{N_{H5}} \int_{\mathbb{R}^{1,5}}  d\tilde{b}^{(2)}_s \wedge \star_6 d\tilde{b}^{(2)}_s 
\nonumber\\
&& + \frac{1}{8 (2\pi) \ell_s^2}  \sum_{s=1}^{N_{H5}} \int_{\mathbb{R}^{1,5}}  \left(-a_1 \, \tr F^2_1 - a_2 \, \tr F^2_2
+ b \, \tr R^2 \right) \wedge \ \tilde{b}^{(2)}_s ,
\nonumber
\eea
and the counter-terms from $\tilde{b}^{(2)}_s \sim \tilde{b}^{(2)}_s$ exchange sum up to
\bea
{\cal I}_{H5,1} =  \frac{N_{H5}}{128 (2\pi)^2 \ell_s^4}  \left( \left[a_1\tr F^2_1 + a_2\tr F^2_2\right]^2 
- 2 b \, \tr R^2  \left(a_1 \tr F^2_1 + a_2 \tr F^2_2\right) 
+ b^2 (\tr R^2)^2   \right).
\eea
We make furthermore the ansatz for a Chern-Simons like coupling of the form\footnote{The
gravitational coupling in~(\ref{E:E8_H5_CS}) was found in~\cite{Witten:1996hc,Becker:1999kh} in the ten-dimensional
set-up by reduction from M-theory.}
\bea
S_{CS,H5}^{E_8 \times E_8} &=& - \frac{2\pi}{\ell_s^6}  \sum_{s=1}^{N_{H5}} \int_{\mathbb{R}^{1,5}} B^{(6)} 
\label{E:E8_H5_CS}
\\
&&+\frac{1}{96 \pi \ell_s^2}\sum_{s=1}^{N_{H5}} 
\int_{\mathbb{R}_{1,5}} B^{(2)} \wedge \left[ \eta_0 \tr R^2 + \eta_1 \tr F_1^2 + \eta_2 \tr F_2^2 \right], 
\nonumber
\eea
where the first term enters the tadpole cancellation condition and the terms in the second line
contribute to the generalised Green-Schwarz mechanism via counter-terms involving the exchange of $b^{(2)}_0 \sim c^{(2)}_0$,
\bea
{\cal I}_{H5,2} &=& \frac{N_{H5}}{192 (2\pi)^2 \ell_s^4} 
\Bigl[\eta_1 (\tr F_1^2)^2 + \eta_2 (\tr F_2^2)^2 + (\eta_1 + \eta_2) (\tr F_1^2)(\tr F_2^2) 
\nonumber\\
&&\quad\quad\quad\quad\quad
+\tr R^2 \left[(\eta_0 -\eta_1)\tr F_1^2 + (\eta_0 -\eta_2)\tr F_2^2 \right] - \eta_0 (\tr R^2)^2
\Bigr].
\eea
Together, we obtain 
\bea
{\cal I}_{H5} &=&  {\cal I}_{H5,1}+ {\cal I}_{H5,2}
\\
&=& \frac{N_{H5}}{192(2\pi)^2 \ell_s^4}  
\Bigl[
\sum_{i=1}^2 \alpha_i (\tr F_i^2)^2 
+\beta \, (\tr F_1^2)(\tr F_2^2)
+\sum_{i=1}^2 \gamma_i (\tr F_i^2)(\tr R^2)
+ \delta \, (\tr R^2)^2
\Bigr]
\nonumber
\eea
with
\bea
 \alpha_i \equiv \frac{3}{2} a_i^2 + \eta_i &\stackrel{!}{=}& 2,
\quad\quad
\beta  \equiv 3  a_1a_2 +\eta_1+\eta_2  \stackrel{!}{=} -2,
\nonumber\\ 
\gamma_i \equiv-3  b a_i +\eta_0 -\eta_i &\stackrel{!}{=}& -1,
\quad\quad  
\delta \equiv  \frac{3}{2} b^2 -\eta_0  \stackrel{!}{=} \half.
\label{E:H5_constants}
\eea
In particular, we have the relations
\bea  
(b-a_i)^2 \stackrel{!}{=} 1 ,
\quad\quad
(a_1 -a_2)^2 \stackrel{!}{=} 4,
\nonumber
\eea
showing that $a_1 \neq a_2$ is necessary.

The overall counter-term takes the correct form for the most symmetric choice
\bea
a_1=-1, \quad a_2=1, \quad b=0, \quad \eta_0=-\half, \quad \eta_1=\eta_2=\half.
\label{E:H5_constants2}
\eea

The non-perturbative Green-Schwarz counter-term~(\ref{E:E8_non_per_GS}) is thus induced by the following
terms in the H5-brane action,
\bea
S_{kin,H5}^{E_8 \times E_8} &=& 
- \frac{\pi}{\ell_s^4} \sum_{s=1}^{N_{H5}} \int_{\mathbb{R}^{1,5}}\tilde{H}^{(3)}_s \wedge \star_6 \tilde{H}^{(3)}_s,
\nonumber\\
S_{CS,H5}^{E_8 \times E_8} &=& - \frac{2\pi }{\ell_s^6} \sum_{s=1}^{N_{H5}} \int_{\mathbb{R}^{1,5}} B^{(6)} 
\label{E:E8_CS_final}
\\
&&+\frac{1}{192 \pi \ell_s^2} \sum_{s=1}^{N_{H5}}
\int_{\mathbb{R}_{1,5}} B^{(2)} \wedge \left[ \tr F_1^2 + \tr F_2^2 - \tr R^2 \right], 
\nonumber
\eea
with
\bea
\tilde{H}^{(3)}_s &=& d\tilde{b}^{(2)}_s + \frac{\alpha^{\prime}}{8} \left(\omega_{Y,1} - \omega_{Y,2}\right).
\eea
This result agrees with the computation of four-dimensional heterotic gauge anomalies in the presence of H5-branes~\cite{Blumenhagen:2006ux}.


\subsection{Examples with $U(3) \times U(1)$ bundles}
\label{Ss:E8_models_Exs}

As an illustration, we present three models with bundles of type~(\ref{E:E8bundle_2}) which trivially fulfill the K-theory constraint,
one with H5-branes and two without.

Consider one of the $U(3)$ bundles $V_i$ defined in~(\ref{E:Def_V_bundles_ex2}) and the corresponding line bundle $L_i$
with $c_1(V_i) = -c_1(L_i)$ embedded in an $E_8$ factor. The resulting charged spectrum from this $E_8$ factor is given in
table~\ref{Tab:E8_Ex}, and the corresponding DUY condition reads
\bea
\rho_2=\rho_3 \quad {\rm for } \quad  V_1, \quad\quad  {\rm or } \quad\quad \rho_1=\rho_3 \quad {\rm for } \quad V_2.
\eea
\begin{table}[htb]
\renewcommand{\arraystretch}{1.2}
\begin{center}
\begin{tabular}{|c||c|c||c||c|}
\hline
\hline
$SO(10) \times U(1)$  & \# H  & \# V  & $SO(10) \times U(1)$  & \# H \\
\hline \hline
$({\bf 45})_0$ & 0 & 1 & $({\bf 16})_{-1} + c.c.$ & 4 \\
$(\1)_0$ & 20 & 1 & $({\bf 10})_2 + c.c.$ & 6\\
& &  & $(\1)_{-4} + c.c.$ & 14 
\\\hline
\end{tabular}
\end{center}
\caption{Charged spectrum from embedding $U(3) \times U(1)$ in one $E_8$ factor. The $U(1)$ factor is in general massive.}
\label{Tab:E8_Ex}
\end{table}
There are three different obvious ways to satisfy the tadpole cancellation conditions:
\begin{enumerate}
\item
The total bundle 
\bea
V= V_i \oplus L_i ,\quad\quad i=1 \; {\rm or} \;  2
\nonumber
\eea
is embedded in one $E_8$ factor, the resulting gauge group is 
\mbox{$SO(10) \times U(1) \times E_8'$}, and twelve H5-branes are needed in order to fulfill the tadpole cancellation condition.
The low-energy spectrum consists of a copy of the states in table~\ref{Tab:E8_Ex}, the vector in the adjoint of $E_8'$,
twelve tensor and hyper multiplets from the H5-branes and the universally present tensor, twenty neutral hyper and the supergravity 
multiplet.
The Abelian gauge factor becomes massive. Fittingly the DUY condition gives one constraint
on the K\"ahler moduli.
\item
The total bundle is 
\bea
V= (V_1 \oplus L_1) \oplus   (V_2 \oplus L_2) 
\nonumber
\eea
and the resulting gauge group $\left[SO(10) \times U(1) \right] \times  \left[SO(10)^{\prime} \times U(1)^{\prime} \right]$.
 The tadpole
cancellation condition is satisfied without H5-branes. $c_1(V_1)$ and $c_1(V_2)$ are linearly independent
leading to two massive Abelian gauge factors. Compatible with this fact, the DUY conditions freeze two K\"ahler moduli.
\item
The total bundle contains two copies of the same vector bundle,
\bea
V= (V_i \oplus L_i)^{\oplus 2},\quad\quad i=1 \;  {\rm or} \;  2,
\nonumber
\eea
resulting again in the  gauge group $\left[SO(10) \times U(1) \right] \times  \left[SO(10)^{\prime} \times U(1)^{\prime} \right]$
and tadpole cancellation without H5-branes. In this case, the linear combination $U(1) - U(1)^{\prime}$ remains massless
while the orthogonal combination acquires a mass. As expected, the DUY condition freezes only one K\"ahler modulus.
\end{enumerate}
By comparing examples 2 and 3, it is obvious that the second Chern characters are not sufficient to describe bundles.
$\ch_2(V_1)=\ch_2(V_2)=-10$ are identical in these examples, but $c_1(V_1) \neq \pm c_1(V_2)$ determines the number of 
massive Abelian gauge factors.


\section{Conclusions}
\label{S_con}

In this article, the six-dimensional generalized Green-Schwarz mechanism for $K3$ compactifications of the
heterotic $SO(32)$ and $E_8 \times E_8$ string with arbitrary Abelian and non-Abelian bundles and five-branes has been derived.
General classes of embeddings have been introduced and their anomaly-eight forms computed.
The dimensional reduction of the ten-dimensional tree-level and one-loop counter-terms matches these anomaly eight-forms
in the absence of H5-branes. For the $SO(32)$ string, the Chern-Simons couplings of H5-branes introduced in~\cite{Blumenhagen:2005zg}  
serve to cancel all remaining six-dimensional field theory anomalies.
For the H5-brane action in the \mbox{$E_8 \times E_8$} heterotic theory, the kinetic terms of the additional tensor 
multiplets~(\ref{E:E8_H5_kin}) together with some Chern-Simons-like coupling~(\ref{E:E8_H5_CS}) provide the correct 
Green-Schwarz counter-terms.

In contrast to the four-dimensional case, the six-dimensional theory admits two different types of Green-Schwarz diagrams 
depicted in figure~\ref{Fig:GS}:
tensors are needed to cancel anomalies involving only non-Abelian gauge fields and gravity, while for Abelian gauge fields
also four-forms and their scalar duals contribute to the anomaly cancellation. The linear couplings~(\ref{E:U1mass}) 
to the four-forms render Abelian gauge fields massive, and  the corresponding Donaldson-Uhlenbeck-Yau and holomorphicity
conditions 
freeze three geometric moduli from the same hyper multiplet. 
The $U(1)$ masses depend on the first Chern classes of the respective bundles times some combinatorial factors
specifying the embedding in $SO(32)$ or $E_8$.

The six-dimensional results are in full agreement with the four-dimensional observations on multiple anomalous $U(1)$ factors
in heterotic compactifications apart from the difference that the DUY condition (most likely) does not receive any loop corrections 
in the present case as argued in section~\ref{Ss_diverse}.

The classification of $E_8$ breakings with one $U(1)$ gauge factor in section~\ref{Ss_E8_models} as well as
more general products of several $U(n)$ bundles in one $E_8$ factor as in~\cite{Blumenhagen:2005ga}
give rise to a very large class of $E_8 \times E_8$ string vacua with multiple $U(1)$ factors. 
Together with the class of $SO(32)$ models in section~\ref{Ss_SO32_H5tcc}, this leads to many models
beyond the classification of six dimensional ${\cal N}=1$ supergravity theories in~\cite{Avramis:2005hc}.

The results are also relevant in order to understand better the F-theory and S-duality relations with 
orientifold compactifications to six dimensions, for which
the analogous dimensional field theory
reduction to six dimensions could be performed along the lines of~\cite{Blumenhagen:2005zh}.

It might be interesting to compactify the heterotic vacua presented here on an additional two-torus, 
and investigate further in the four-dimensional ${\cal N}=2$ field theory set-up
T-duality among the two heterotic theories as well as the relation to type II 
Calabi-Yau compactifications.

Finally, the toy examples presented in this article contain only up to three $(1,1)$-forms. Fully fledged models 
involving all 22 two-forms could be obtained via the spectral cover construction 
on elliptically fibered $K3$ manifolds~\cite{Bershadsky:1997ec,Friedman:1997yq}.

 \vskip 1cm
 {\noindent  {\large \bf Acknowledgments}}
\vskip 0.5cm 
It is a pleasure to thank Ralph Blumenhagen and Timo Weigand for many valuable discussions and Katrin Wendland for a useful communication.
\vskip 2cm

\clearpage


\end{document}